\documentclass[12pt]{spieman}  
\usepackage{amsmath,amsfonts,amssymb}
\usepackage{graphicx}
\usepackage{setspace}
\usepackage{tocloft}
\usepackage{soul}
\usepackage{lineno}

\title{Learning the Lantern: Neural network applications to broadband photonic lantern modelling}

\author[a, b, *]{David Sweeney}
\author[a, b]{Barnaby R. M. Norris}
\author[a, b]{Peter Tuthill}
\author[c]{Richard Scalzo}
\author[a, b]{Jin Wei}
\author[a, b]{Christopher H. Betters}
\author[a, b]{Sergio G. Leon-Saval}
\affil[a]{The University of Sydney, Faculty of Science, School of Physics, Physics Road, Sydney, Australia, 2006}
\affil[b]{Sydney Astrophotonic Instrumentation Laboratories, Physics Road, Sydney, Australia, 2006}
\affil[c]{The University of Sydney, Faculty of Science, School of Mathematics \& Statistics, Physics Road, Sydney, Australia, 2006}

\cftpagenumbersoff{figure}
\cftpagenumbersoff{table} 
\begin{document} 
\maketitle

\begin{abstract}
Photonic lanterns allow the decomposition of highly multimodal light into a simplified modal basis such as single-moded and/or few-moded. They are increasingly finding uses in astronomy, optics and telecommunications. Calculating propagation through a photonic lantern using traditional algorithms takes $\sim 1$ hour per simulation on a modern CPU. This paper demonstrates that neural networks can bridge the disparate opto-electronic systems, and when trained can achieve a speed-up of over 5 orders of magnitude. We show that this approach can be used to model photonic lanterns with manufacturing defects as well as successfully generalising to polychromatic data. We demonstrate two uses of these neural network models, propagating seeing through the photonic lantern as well as performing global optimisation for purposes such as photonic lantern funnels and photonic lantern nullers.
\end{abstract}


{\noindent \footnotesize\textbf{*}Corresponding author,  \linkable{david.sweeney@sydney.edu.au} }

\begin{spacing}{1.2}   

\section{Introduction}
\label{sec:intro}  

Photonic lanterns (PLs)\cite{Leon-Saval:05} are increasingly important devices finding use in fields such as astronomy, for wavefront sensing\cite{Norris2020AnSensor} and transforming astronomical signals\cite{Nem2020, Diab:19, Anagnos:18}; in optics, for reducing modal noise\cite{Cvetojevic:17}, amplification\cite{Montoya:17} and beam shaping \cite{Gross:19}; and in telecommunications, for free space communication through the atmosphere\cite{Ozdur:13} and spatial division multiplexing\cite{Fontaine:13, Thomson:16, Essiambre:13, LEONSAVAL201746, Eznaveh:18, chandrasekharan:17}. Of particular astronomical interest is the use of PLs to efficiently couple seeing-affected light into single mode fibres for single-mode spectroscopy applications\cite{Bland-Hawthorn2010PIMMS:Microspectrograph, Betters:13, Schwab2012SingleSpectrographs}, including recent on-sky tests\cite{NemSMF, NemStrehl}.  

Hampering this widespread interest, traditional numerical beam propagation algorithms (such as {\tt RSoft}'s BeamPROP) are only capable of calculating propagation through a three dimensional PL in $\sim 1$ hour per simulation on a contemporary quad-core CPU (we state time per quad-core here as a useful metric since virtually all modern CPUs are built around a quad-core design). {These beam propagation algorithms solve Maxwell's equations at each point of a fine grid, interspersed through the medium.} 
This is too slow for many applications, putting the realistic simulation of a PL's response to atmospheric-seeing time sequences well out of reach, particularly for polychromatic data, and prevents the performing of global optimisations, e.g. to find a specific input wavefront for a given output criteria. In this latter case, basin-hopping gradient descent (or some other optimisation algorithm) can be deployed to use the PL for novel applications like PL funnels and PL nullers.
The inability of traditional beam propagation algorithms to rapidly simulate PLs motivates our use of neural networks (NN) as a faster alternative.

{A fundamental challenge is that the relationship between the wavefront at a PL's input and the measured intensities at its output is non-linear (since the intensity is the square of the complex electric field). NNs are thus extremely valuable to model this since they can be truly non-linear. Across many fields,} NNs are rising to prominence for their ability to approximate arbitrary{, non-linear} functions. 
Buoyed by the large data sets and increasing computing power which characterises modern machine learning problems, NNs now outperform many classic approaches for a wide variety of problems \cite{Krizhevsky2012ImageNetNetworks, Silver2016MasteringSearch, Ganesan2010ApplicationData, Wei2016NeuralReactions}.

Here, we explore the deployment of deep learning to synthesise these elements, achieving a speed-up compared to traditional beam propagation algorithm of greater than 5 orders of magnitude. 
{Another key advantage of this approach is that the model can be built from in-situ test measurements (`training data') of a real PL. These measurements include the effects caused by the manufacturing defects and an imperfect optical injection system, rather than the idealised parameters used to create an RSoft model. Alternatively the training data can be initially produced from test-propagations through a conventional beam-propagation algorithm. Moreover, it generalises to broadband data and is able to deliver accurate predictions for untrained wavelengths.
}
We demonstrate the case of starlight distorted by natural astronomical seeing passing through a PL, with only tens of milliseconds of compute time per simulation. 
Furthermore, two applications discussed below that would formerly have required substantial resources are now straightforward with this new capability.

\subsection{Photonic Lanterns}

\begin{figure}[htb!]
    \centering
    \includegraphics[width=1.0\textwidth]{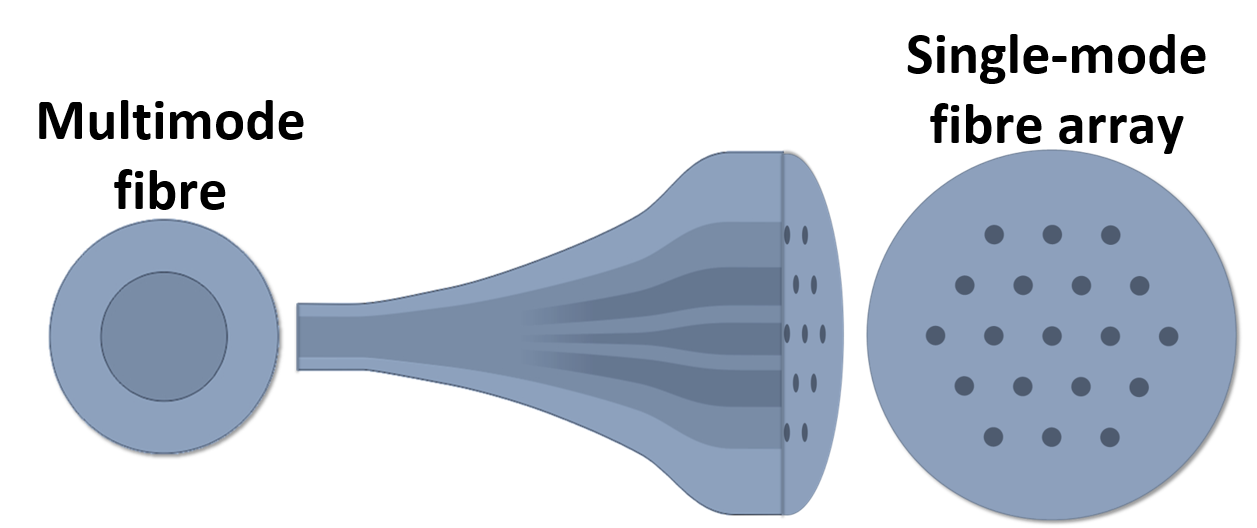}
    \caption[Photonic Lantern Diagram]{A diagram of a PL depicting the splitting of modes from one multimode optical fibre to a number of individual single-mode optical fibres.}
    \label{fig:photonic_lantern}
\end{figure}{}

{\em PLs} are composed of a multimode optical fibre section which adiabatically transitions into many single-mode optical fibres, as shown in Fig.~\ref{fig:photonic_lantern}.
Light incident on the multimode optical fibre propagates through the multimodal region before gradually transitioning into an array of single-mode fibres---with higher refractive index---which increase in diameter in this direction of propagation. 
Light propagating through the allowed modes of the lantern is thus distributed, in varying fluxes, among the array of single-mode fibres \cite{Leon-Saval2013PhotonicLanterns}.  In order to conserve flux, the number of outputs must equal the number of modes in the multimode optical fibre.
The phase and intensity of the output of each single-mode fibre encodes the information present in the original multimode field, up to the complexity allowed by the modes supported by the multimode region\cite{Norris2020AnSensor}.
If a suitable model of the PL existed, then by measuring the incident multimode field it would be possible to determine the output intensities of each single-mode fibre.

Calculating the intensities of the array of single-mode fibres to any given multi-mode input field is non-trivial. For an analytic solution of a physical PL, the coupling coefficients must be determined experimentally---they cannot be precisely specified during the design due to imperfections in the fabrication process.
For a photonic device, such as a 19 core hexagonally packed PL each with a changing cross section along the length, determination of the coupling coefficients is not straightforward---even using coupled-mode theory with nearest neighbour approximations. 
We demonstrate, in Sec.~\ref{sec:model_performance} that machine learning techniques can be formulated into a system able to learn and predict a lantern's transfer function when trained on the lantern's response to a set of randomised phase input fields. This technique is successful across a wide range of input wavefronts, described here by the Zernike polynomials.

{It should be noted that the quantity measured at the output of a PL in an astronomical setting (either through a spectrograph or photo detector) is the waveguide intensity. Since this is the square of the complex electric field, its relationship with the input phase or amplitude is non-linear (making the use of a NN desirable). Furthermore, it is degenerate---a particular set of intensities at the single-mode outputs can map to multiple combinations of phase and amplitude at the multimode input.}

\subsection{Neural Networks}

A NN is a machine learning technique for approximating complicated functions. 
The power of this approach is that by confronting a NN with a data set the network can ``learn" a mapping through a process called backpropagation \cite{Rumelhart1986LearningErrors}. Once a mapping has been learned the NN can be used to predict a point in the output space from the corresponding point in the input space---allowing a fast conversion between the two spaces. This learning process happens largely without human input and so high performance computers can be employed to determine a mapping for tasks which are too tedious, too complex or too fast for humans.

NNs are {specified} by a number of {\em hyperparameters} which describe the networks complexity (such as the number of nodes) and regularisation ({a school of techniques for reducing overfitting,} for example ``dropout''\cite{srivastava:14}). As the design of a NN iterates, hyperparameters are tuned to establish values which are optimised for the problem at hand. A review of the variety of hyperparameters and an in-depth explanation of NNs can be found in  Ref.~\citenum{10.3389/frai.2020.00004}, or a shorter review can be found in Ref.~\citenum{Lecun2015}.

In Sec.~\ref{sec:simulated_data} we describe the training of NN models based on simulated data created in RSoft's BeamPROP. In Sec.~\ref{sec:lab_data} we use the same approach to train NN models on optical testbed data taken using an existing PL in both monochromatic and polychromatic light. We then describe the performance of these trained models in Sec.~\ref{sec:model_performance} before demonstrating two experiments performed with these models in Sec.~\ref{sec:model_applications}.

\section{Creating a neural network model using simulated data}
\label{sec:simulated_data}
In order to simulate the effects of wavefronts distorted by astronomical seeing passing through a PL or to identify wavefronts which maximise or minimise the output of a single-mode fibre, we require a data set with which to train the NN model. In this work we explore training these NN models on simulated data and on lab measurements of an existing PL, which are compared in Sec.~\ref{sec:model_performance}. 
Regardless of the nature of the data used, more accurate NN models can be trained if larger data sets are available. 

When modelling a theoretical lantern, training data is first produced using a conventional beam propagation algorithm. {Note that the NNs approximate the outcomes of the beam propagation algorithm, rather than explicitly solving Maxwell's equations. Thus, the NNs do not replace traditional beam propagation algorithms, which are required to create a training set which captures the physics at play in this system.} The creation of this training data set takes some time, but once produced the resulting NN model can be used to rapidly calculate the transmission of any arbitrary wavefront through the PL. {The time taken to create this training data is explored at the end of this section.}

The simulated data used in this work was generated through the {\em BeamPROP} module of RSoft, a three dimensional optical modelling software that numerically propagates electromagnetic waves through media of almost arbitrary shape and properties. 
This ability to handle micro-patterned structures was exploited to simulate the functioning of a three dimensional PL. We modelled a 19-output multicore fibre lantern with physical properties summarised in Table~\ref{table:RSoft_PL_properties}. Propagation in BeamPROP for a 3D PL with a sufficiently large volume and fine grid size requires about 60 minutes per simulation {on four cores} of an AMD Threadripper 2990WX. 

\begin{table}[htb!]
\caption[Photonic Lantern Properties]{The physical properties of the PL, modelled in RSoft, used in this project.}
\label{table:RSoft_PL_properties}
\begin{center}
\begin{tabular}{|| c | c ||} 
 \hline
 \shortstack{Property} & \shortstack{Value}\\ [0.5ex] 
 \hline\hline
 Single-mode waveguides & 19 \\ 
 Waveguide packing & Centred hexagonal \\
 Waveguide length & 40\,000~$\mu m$ \\
 Waveguide taper & Linear \\
 Waveguide final width, height & 6.5~$\mu m$\\
 Waveguide refractive index & 1.447\\
 Cladding refractive index & 1.026\\
 Background refractive index & 1 \\[1ex]
 \hline
\end{tabular}
\end{center}
\end{table}

The incident light field was assumed to have uniform intensity in the pupil plane, with a randomly varying phase term formed by the superposition of Zernike polynomials in Noll order\cite{Noll:76}.
The strength of each individual Zernike term is governed by a coefficient specifying the {root mean squared} (RMS) wavefront error (WFE) contributed by that term. Data sets were created in which these coefficients were drawn from {one of three ranges:} {$\pm 0.85 $~radians (}$\pm 0.2$~$\mu m${)}, {$\pm 2.1$~radians (}$\pm 0.5$~$\mu m${)} and {$\pm 4.1$~radians (}$\pm 1$~$\mu m${)}. {With one of the ranges selected, coefficients were drawn independently and uniformly from within the bounds of the range. Larger ranges, such as $\pm 2.1$~radians and $\pm 4.1$~radians we used to ascertain the model's performance on wavefronts with a larger RMS WFE. This uniform sampling is likely not optimal in terms of sample efficiency; however, we chose this method to avoid biasing our data.} Poppy \cite{Perrin2012SimulatingWebbPSF} was used to generate wavefronts described by these Zernike polynomials{. Once created these wavefronts were focused, by Poppy, as if} through an ideal lens and into {point spread functions (PSFs)}{. These PSFs} were then passed to RSoft to be injected into the simulated PL{. T}he intensities of the single-mode waveguides were {then extracted from RSoft, and recorded.}

The set of Zernike polynomial coefficients used to construct the wavefront and their corresponding set of waveguide intensities formed one training example. Thousands (see Table~\ref{table:models_simulated}) of these examples are combined to create the {\it training, validation} and {\it testing} sets. Since the model has seen the training data (during training) we assess the model's performance on the previously unseen validation set to tune the hyperparameters. As hyperparameters are tuned to optimised performance on the validation set, we can inadvertently bias the model's performance on this set. For this reason, the unseen testing set is used to provide the final metrics. In order to ensure the testing set was large enough to provide accurate metrics each simulated data set has a testing set of 1\,000 examples. 

We can exploit a symmetry in the PL to augment our training set, i.e. increase the size of the set without additional computational power.
The PL used in this data set has its single-mode fibre waveguides arrayed in a centred hexagonal lattice which has six-fold rotational symmetry along each interval of $\frac{\pi}{3}$. 
By rotating our input PSF by $\frac{\pi}{3}$ radians {the pattern observed in the output waveguides will also rotate} by the same amount {in an ideal PL. Due to manufacturing defects this does not currently work on real-world PLs.} 
In order to rotate the PSF, the Zernike polynomials are rotated by adjusting the weight given to each polynomial based on Eq.~(\ref{eq:12}) of Ref.~\citenum{Li2018AnalyticalPupils}:
\begin{align}
    \Vec{c'} &= \Vec{T}\Vec{c} \label{eq:12}\\
    \Vec{T}_{nmn'm'} &= \delta_{nn'}\delta_{|m||m'|}\begin{cases} 
      1 & m' = 0 \\
      \cos (m'\phi) & m' = m\\
      \sin (m'\phi) & m' = -m
   \end{cases} \label{eq:13}
\end{align}
Where $\Vec{c'}$ is the vector of Zernike coefficients in the rotated frame, $\Vec{c}$ is the vector of Zernike coefficients in the original frame and the elements of $\Vec{T}$ are defined in Eq.~(\ref{eq:13}). In Eq.~(\ref{eq:13}), $\phi$ is the angle by which $\Vec{c'}$ will be rotated, $n$ and $n'$ are the radial degree of each Zernike polynomial in their respective frames and $m$ and $m'$ are the azimuthal frequency of each Zernike polynomial in their respective frames. 

RSoft increments the propagating wavefront through the PL with a user-set interval: the grid size parameter. 
There is a balance to be struck with this grid size parameter between fidelity and speed, as a finer grid size will produce more accurate simulations at the cost of a longer compute run. 
This trade space is explored in Table~\ref{table:grid_sizes}, which illustrates the strong computational penalty imposed by finely sampled spatial grids.
The utility of the lower-fidelity simulations is examined later in this work. 

\begin{table}[htb!]
\caption[Grid size run-times]{Comparison of mean simulation rate for varying RSoft grid sizes. These simulations were run on {4 CPU cores} ({on} an AMD Threadripper 2990WX) with 16 simulations running in parallel.}
\label{table:grid_sizes}
\begin{center}
\begin{tabular}{||c | c | c | c ||} 
 \hline
 \shortstack{X/Y Grid Size\\($\mu m$)} & \shortstack{Z Grid Size\\($\mu m$)} & \shortstack{Number of Simulations \\in Test Run} & \shortstack{{Mean Minutes}\\ {per Simulation}}\\ [0.5ex] 
 \hline\hline
 0.25 & {2} & 1\,000 & {51.2} \\ 
 0.5 & {2} & 4\,000 & {13.0} \\ 
 1 & {2} & 16\,000 & {3.84} \\[1ex] 
 \hline
\end{tabular}
\end{center}
\end{table}

\section{Creating a model using photonic lantern data from an optical laboratory testbed}
\label{sec:lab_data}
Two sets of data were taken in the laboratory using monochromatic and polychromatic illumination. 
The monochromatic configuration employed a $685$~nm laser with a measured bandwidth of $1.2$~nm, collimated to create a plane wave. The beam was then shaped to deliver the required diverse population of input wavefronts using a spatial-light modulator. Each pattern was determined by drawing a set of Zernike polynomial coefficients from a uniform distribution, each yielding a value between approximately -0.4 and 0.4~radians RMS WFE. 
The monochromatic data set used only the first 10 Zernike polynomials, resulting in a mean RMS WFE of $0.744$~radians. 

The polychromatic data set employed the same setup, replacing the laser with a broadband halogen light source.
Furthermore{,} here the first 20 Zernike polynomials in OSA\cite{Thibos:00} order (corresponding to the first 21 Zernike polynomials in Noll order, excluding the 20th) were employed 
yielding a mean RMS WFE of $0.950$~radians.

The wavefront was then injected into the multimode region of the PL with a microscope objective. The PL had a visible wavelength multicore fibre with a diameter of 22~$\mu m$ and a numerical aperture of 0.145. The multimode region transitions to 19 single-mode cores each with a diameter of 3.7~$\mu m$, a numerical aperture of 0.14 and a core-to-core separation of 35~$\mu m$. The outputs of the single-mode cores were then imaged by a camera with an $f=200$~$mm$ doublet lens. 
Images collected with this arrangement were accumulated over thousands of input wavefronts to create a data set, as summarised in Sec.~\ref{sec:model_performance}. The process is extremely rapid: the rate at which data can be produced is limited only {by} the frame rate of the camera and software latency ($\ll$ 1 second per wavefront). {Note that this approach treats the entire experimental set-up as one system which will be modelled by a NN. As such we do not need to model every single lens, but will require that we re-characterise the system if the PL is moved to a different set-up.}

For the polychromatic data, the multicore output was collimated and spectrally dispersed before imaging onto the camera, with the resulting spectra yielding 31 wavelength channels spaced between $655.8$--$751.5$~nm for each core. 

The monochromatic lab data set had a testing set containing 1\,163 examples. The broadband lab data set had a testing set size of 31\,000 examples comprised of Zernike coefficient combinations which the network had not seen.

\section{Results and Discussion}
\subsection{Model Performance}
\label{sec:model_performance}
We trained densely connected feed-forward NNs using Keras \cite{Chollet2015Keras} with an Adam \cite{Kingma2015Adam:Optimization} optimiser. We used mean squared error (MSE) as our loss function, Leaky ReLU as our activation function and batch normalisation \cite{Ioffe2015BatchShift} was used between hidden layers.

We sought to keep the statistical variation in the validation and testing sets as close to a real-world distribution as possible, so that recorded metrics were directly comparable to realistic applications. To this end, when using simulated data only the highest fidelity data were included in the validation and testing sets and no augmentation was applied. The lower fidelity data (generated with grid sizes $> 0.25$~$\mu$m) was assigned entirely to the training set and this was the only set on which data augmentation was applied.

\begin{figure}[htb!]
    \centering
        \includegraphics[width=1\textwidth]{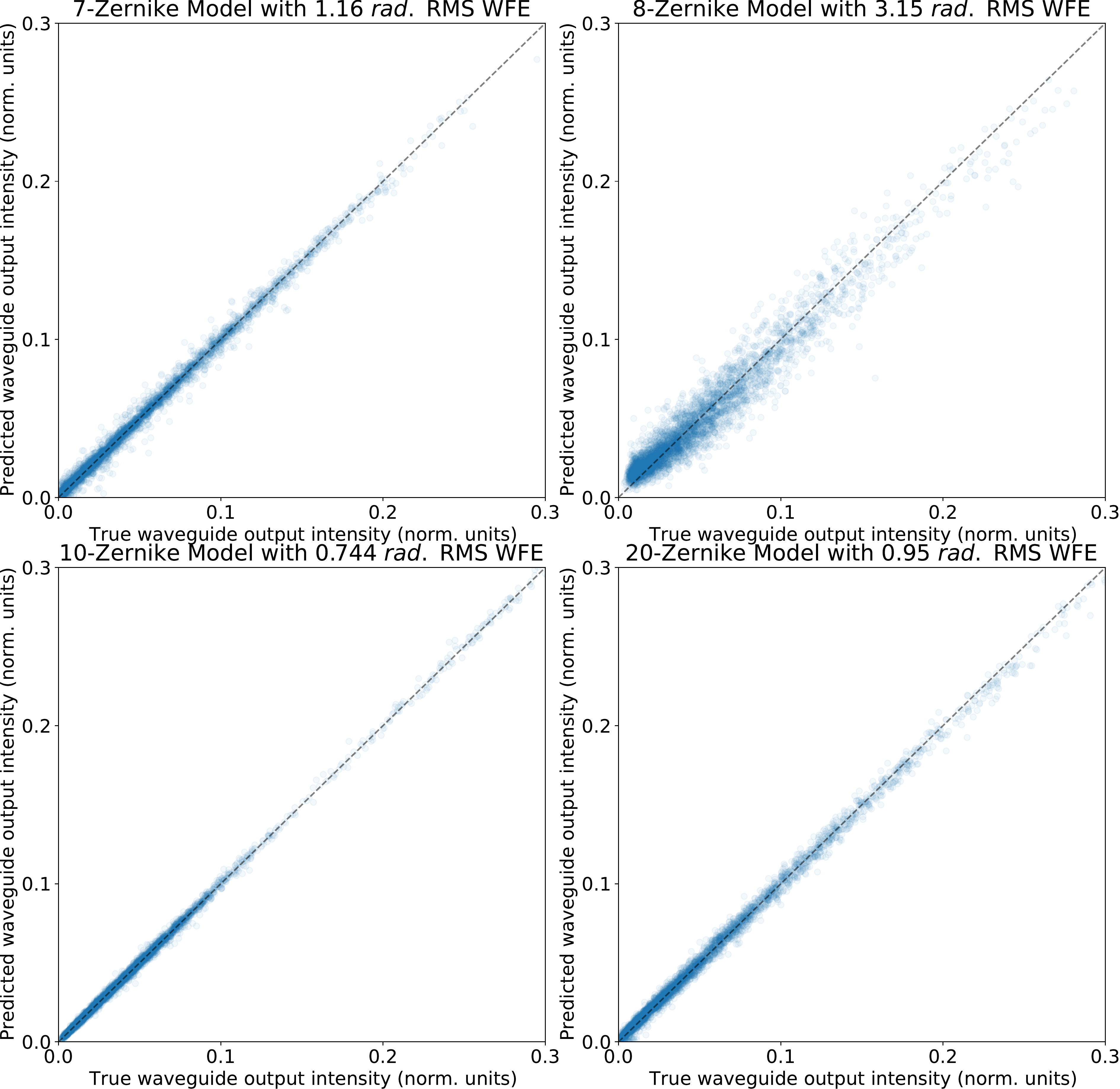}
    \caption[Correlation plot of model predictions]{Correlation plot of a random selection of examples from each NN model's testing set. The top row of plots are from the best performing models trained on simulated data (Table~\ref{table:models_simulated}). The bottom row are from models trained on laboratory measurements of an existing PL (Table~\ref{table:models_practical}), i.e. the bottom-left plot is from the monochromatic model and the bottom-right plot is from the polychromatic model. The black dashed line denotes perfect predictions. The outputs of the model were normalised such that, on average, sets of intensities from the data set summed to 1.}
    \label{fig:test_predictions}
\end{figure}

Hyperparameters were iteratively tuned to optimise the loss function on each data set. The difference in MSE between an untuned NN model and one which had been tuned (i.e. {hyperparameters} adjusted from some initial value) tended to be around a factor of 2. Ultimately, densely connected feed-forward networks with hidden layers of constant sizes were found to perform best. A list of these hyperparameters and the MSE of each model is presented in Table~\ref{table:models_simulated} for the NN models trained on simulated data sets and Table~\ref{table:models_practical} for NN models trained on the lab data sets. These tables show that {increased RMS WFE and the number of Zernike polynomials requires more complex models (increased layers) and also results in larger MSE}. Plots depicting some of these models can be seen in Fig.~\ref{fig:test_predictions}. 

\begin{figure}[htb!]
    \centering
        \includegraphics[width=1\textwidth]{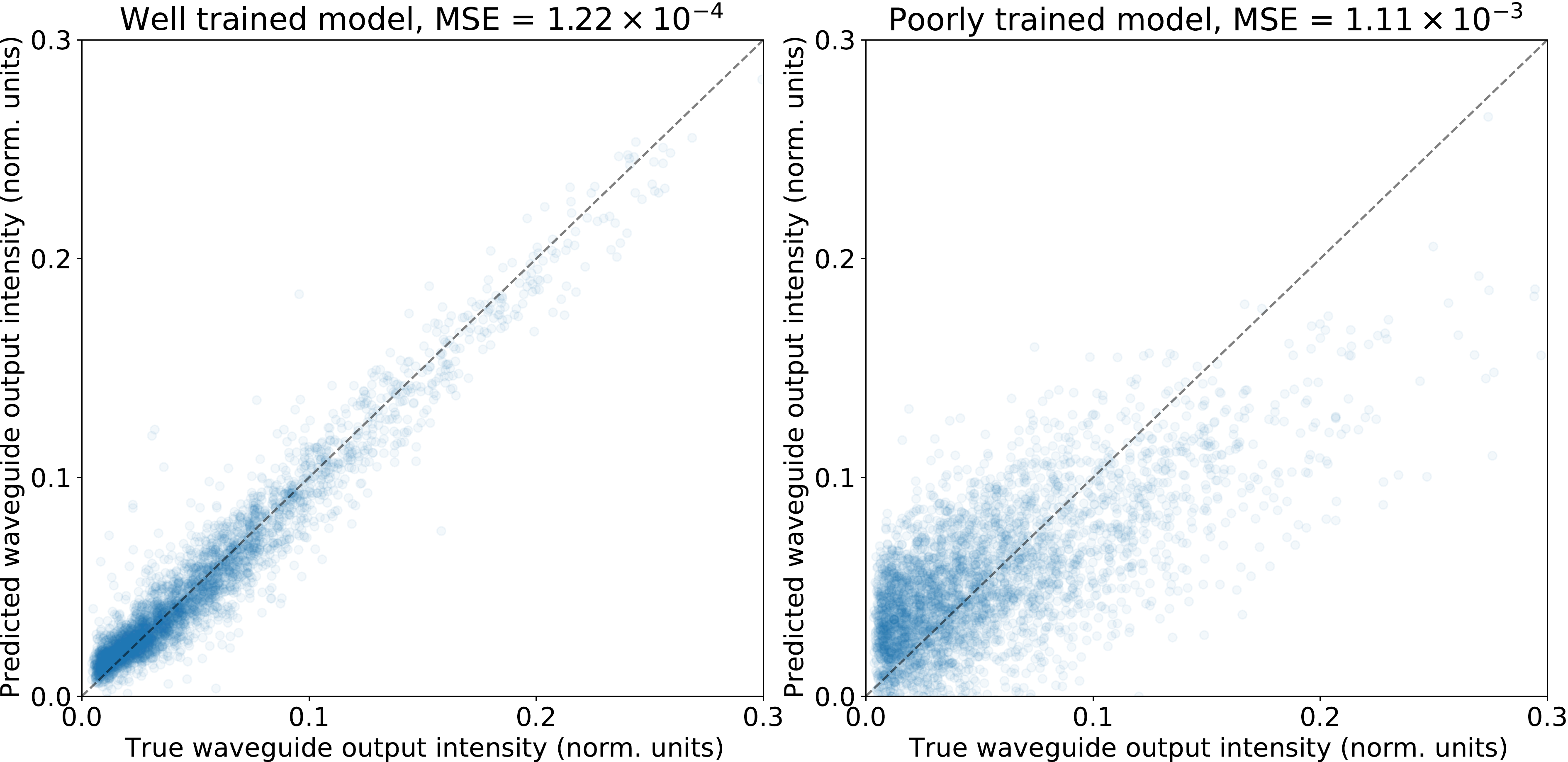}
    \caption[Good versus poor MSE]{{Correlation plot showing the difference between a model with low MSE (the left panel) and a model which was intentionally trained poorly to have a high MSE (the right panel). The poorly trained model is under-regularised and has overfit the training data.}}
    \label{fig:low_mse_comparison}
\end{figure}

{With RMS WFE greater than 1 radian the output intensities are not a linear function of input phase}\cite{Norris2020AnSensor}. The low MSEs throughout demonstrate that not only is this approach capable of emulating simulated PLs but that the approach also produces extremely accurate results on real-world PLs {and also that the models are capable of accurately predicting even when the RMS WFE is very large}. {Figure~}\ref{fig:low_mse_comparison}{ depicts the difference between a low MSE and an (intentionally) high MSE.}

\begin{table}[htb!]
\caption[Best models on simulated data]{Summary of the NN models which best emulated a simulated PL given restrictions on the number and magnitude of the Zernike polynomial coefficients. $N_z$ is the number of Zernike polynomials which describe the input wavefront, $Reg.$ is the method of regularisation used and $p$ is the regularisation hyperparameter. The outputs of the model were normalised such that, on average, sets of intensities from the data set summed to 1. The number of training examples in brackets is the number of training examples before six-fold augmentation through rotation. Note that these bracketed numbers are largely comprised of the data with a coarse grid size.}
\begin{center}
\label{table:models_simulated}
\begin{tabular}{|| c | c | c | c | c | c | r | r ||} 
 \hline
 \shortstack{Mean\\ RMS WFE\\ ({$radians$})} & \shortstack{$N_z$} & \shortstack{Model\\ MSE\\ (norm. units)} & \shortstack{Layers} & \shortstack{Hidden\\ Layer\\ Size}  & \shortstack{$Reg.$} & \shortstack{$p$} & \shortstack{Training\\ Examples}\\ [0.5ex] 
 \hline\hline
 {$1.16$} & 7 & 2.12$\times 10^{-5}$ & 3 & 5\,000 & L2 & 1.58$\times 10^{-7}$ & 1\,672\\
 {$1.52$} & 11 & 2.26$\times 10^{-4}$ & 4 & 5\,000 & L2 & 1.58$\times 10^{-7}$ & 1\,672\\
 {$1.86$} & 16 & 7.00$\times 10^{-4}$ & 4 & 2\,000 & L2 & 3.98$\times 10^{-6}$ & 1\,900\\
 {$2.04$} & 19 & 7.91$\times 10^{-4}$ & 17 & 2\,000 & None & - & 6\,031 \\
 {$3.15$} & 8 & 1.22$\times 10^{-4}$ & 13 & 2\,000 & Dropout & 0.2 & 93\,528 (15\,588) \\
 {$3.77$} & 11 & 4.38$\times 10^{-4}$ & 9 & 2\,000 & Dropout & 0.2 & 291\,714 (48\,619)\\
 {$6.28$} & 8 & 8.42$\times 10^{-4}$ & 15 & 2\,000 & Dropout & 0.2 & 194\,334 (32\,389)\\
 \hline
\end{tabular}
\end{center}
\end{table}

\begin{table}[htb!]
\caption[Best models on lab data]{Summary of the NN models which best emulated a physical, pre-existing PL given restrictions on the number and magnitude of the Zernike polynomial coefficients. The 20-Zernike model was trained and assessed on broadband data, it predicts outputs based on the first 20-Zernike polynomials in ANSI order and the wavelength of light specified. Each Zernike polynomial coefficient was selected from a uniform distribution, as specified in the table. $N_z$ is the number of Zernike polynomials which describe the input wavefront, $Reg.$ is the method of regularisation used and $p$ is the regularisation hyperparameter. The outputs of the model were normalised such that, on average, sets of intensities from the data set summed to 1. 
These data were not augmented through rotation.}
\label{table:models_practical}
\begin{center}
\begin{tabular}{|| c | c | c | c | c | c | r | r ||} 
 \hline
 \shortstack{$N_z$} & \shortstack{Mean RMS \\WFE ($radians$)}  & \shortstack{Model\\ MSE\\ (norm. units)} & \shortstack{Layers} & \shortstack{Hidden\\ Layer\\ Size}  & \shortstack{$Reg.$} & \shortstack{$p$} & \shortstack{Training\\ Examples}\\ [0.5ex] 
 \hline\hline
 10 & $0.744$ & 2.97$\times 10^{-6}$ & 2 & 5\,000 & L2 & 1.58$\times 10^{-7}$ & 57\,570\\
 20* & $0.950$ & 1.90$\times 10^{-5}$ & 8 & 2\,000 & Dropout & 0.2 & 1\,147\,000 \\
 \hline
\end{tabular}
\end{center}
\end{table}

A single high-resolution propagation in BeamPROP, using a single {quad-core} CPU takes around 60 minutes. By using multiple cores per propagation, running propagations in parallel, using a coarser grid size and augmenting the data through rotations we are able to reduce the time required to create these training examples. Using these techniques to accelerate data generation, 250\,000 training examples can be generated in a week (compared with almost 2~years if no augmentation or  {higher}-fidelity data is used). 
Imperfections in the optical testbed removed the symmetries required for data augmentation.
However, lab data {can} be taken much more quickly than simulated data and so augmentation is not required in this case.

\begin{figure}[htb!]
    \centering
        \includegraphics[width=0.45\textwidth, height=5.5cm, keepaspectratio]{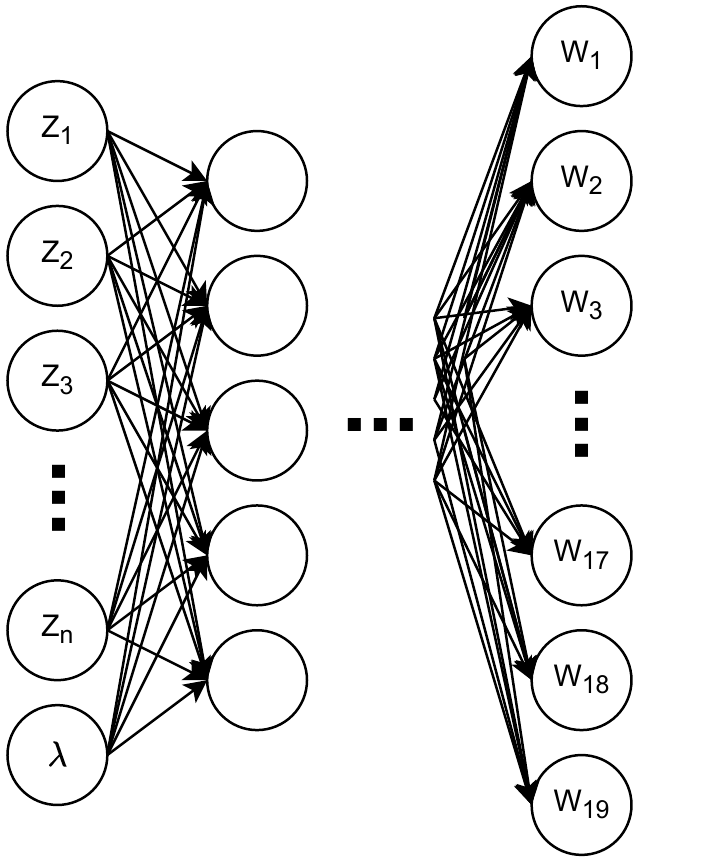}
    \caption[Polychromatic network structures]{Schematics showing the network structure for the polychromatic data. In this figure $Z_n$ is the $n$th Zernike polynomial and $W_n$ is output of the $n$th waveguide. In this structure the wavelength is included as an input in addition to the Zernike polynomial coefficients. 
    }
    \label{fig:broadband_network_structures}
\end{figure}

For the polychromatic data set we used an architecture which included the wavelength as an input to the model, as shown in Fig.~\ref{fig:broadband_network_structures}. This architecture has the added advantage that the network is able to learn how the PL outputs vary with wavelength, making the network capable of generalising to unseen wavelengths.

As discussed in Sec.~\ref{sec:lab_data}, this NN model was trained on data generated by 31 discrete wavelengths ranging from $655.8$--$751.5\,nm$. To demonstrate that the model accurately predicts unseen wavelengths in this range, a new NN model was trained on wavelengths from $655.8$--$697.2\, nm$ and $703.5$--$751.5\, nm$ and tested on $700.3\,nm$ wavelength data. This NN model, which had identical hyperparameters to the model described in Table~\ref{table:models_practical}, achieved a MSE of 1.97$\times 10^{-5}$---practically identical to the predictions made by the full NN model.

\subsection{Model Experiments}
\label{sec:model_applications}

We demonstrate two possible use-cases of the NN models created in Sec.~\ref{sec:model_performance}. Firstly, we show its performance in response to simulated astronomical seeing (or other sets of arbitrary WF error) and secondly it is employed to design optimal injection WFs for specific output-patterns using global optimisation. In this second case, we chose to formulate designs for two potentially valuable photonic processes: {\em PL funnels} and {\em PL nullers}. The advantage of these neural-network models over traditional beam propagation algorithms lie in their speed and low hardware requirements. Propagating light through a 3D PL to establish its outputs at a suitably fine grid-size with RSoft's BeamPROP takes $\sim$60 minutes. In contrast, the NN models described in Sec.~\ref{sec:model_performance} are 5 orders of magnitude faster{. We performed 100\,000 tests and found the mean time to propagate through the NN model was} around 0.03 seconds per wavefront propagation when run on a consumer-grade GPU card (here an Nvidia RTX 2080TI).

\subsubsection{Response to Seeing}
\label{sec:seeing_response}
\begin{figure}[htb!]
    \centering
    \includegraphics[width=0.95\textwidth]{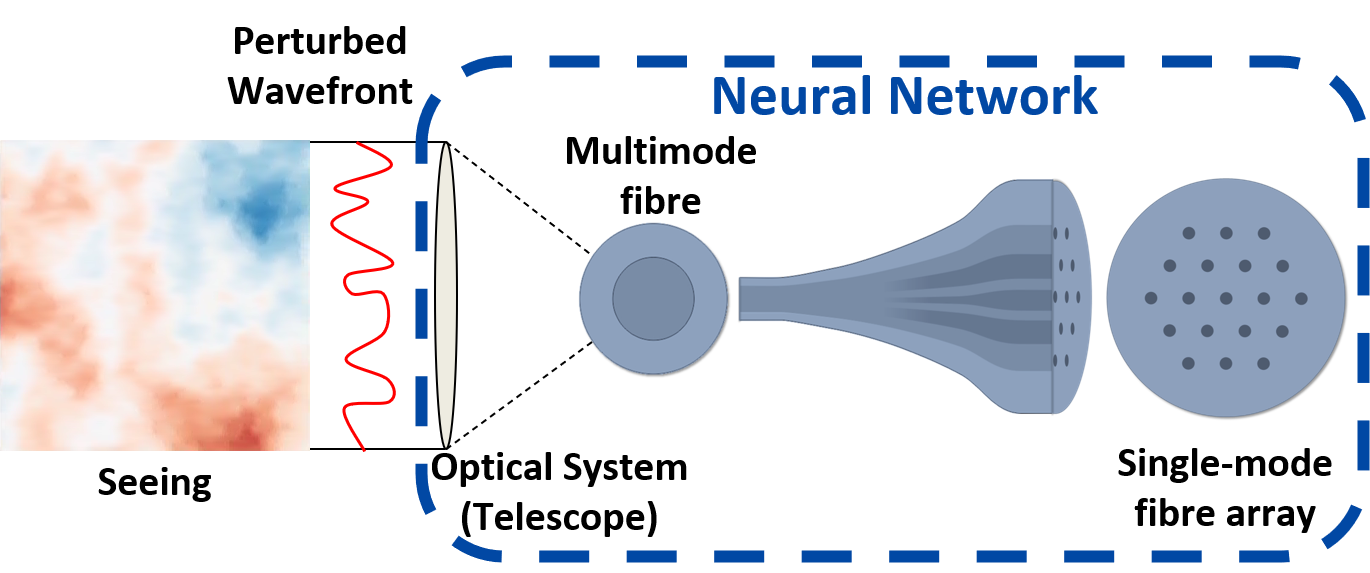}
    \caption[Possible setup whereby seeing is passed into a photonic lantern]{Diagram showing starlight wavefronts perturbed by astronomical seeing and subsequently passed into a PL, resulting in varying outputs. The dashed blue box is the section of the system emulated by the NN models trained in this work.}
    \label{fig:seeing_PL}
\end{figure}

In the design and evaluation of astronomical systems utilising PLs, it is important to have an end-to-end model of the system. However, modelling the response of such a system to realistic seeing has been impractical, with an hour per timestep needed when using beam propagation algorithms. Instead, unrealistic approximations were used---such as flux being evenly distributed between all outputs regardless of WFE\cite{10.1093/mnras/staa3752}. 
Using the NN models described here, this can be done in near real-time.  Fig.~\ref{fig:seeing_PL} depicts a conceptual scheme where a lantern is deployed to measure these effects. 
As illustrated, the seeing results in distorted wavefronts arriving at the telescopes, which then propagate through to a focus at the input of a PL. Note that when data comes from a real PL, such as one in-situ at an adaptive optics (AO) facility, the entire system from deformable mirror (DM) to single-mode outputs can be modelled---not only the PL.

{}

\begin{figure}[htb!]
    \centering
    \includegraphics[width=0.8\linewidth]{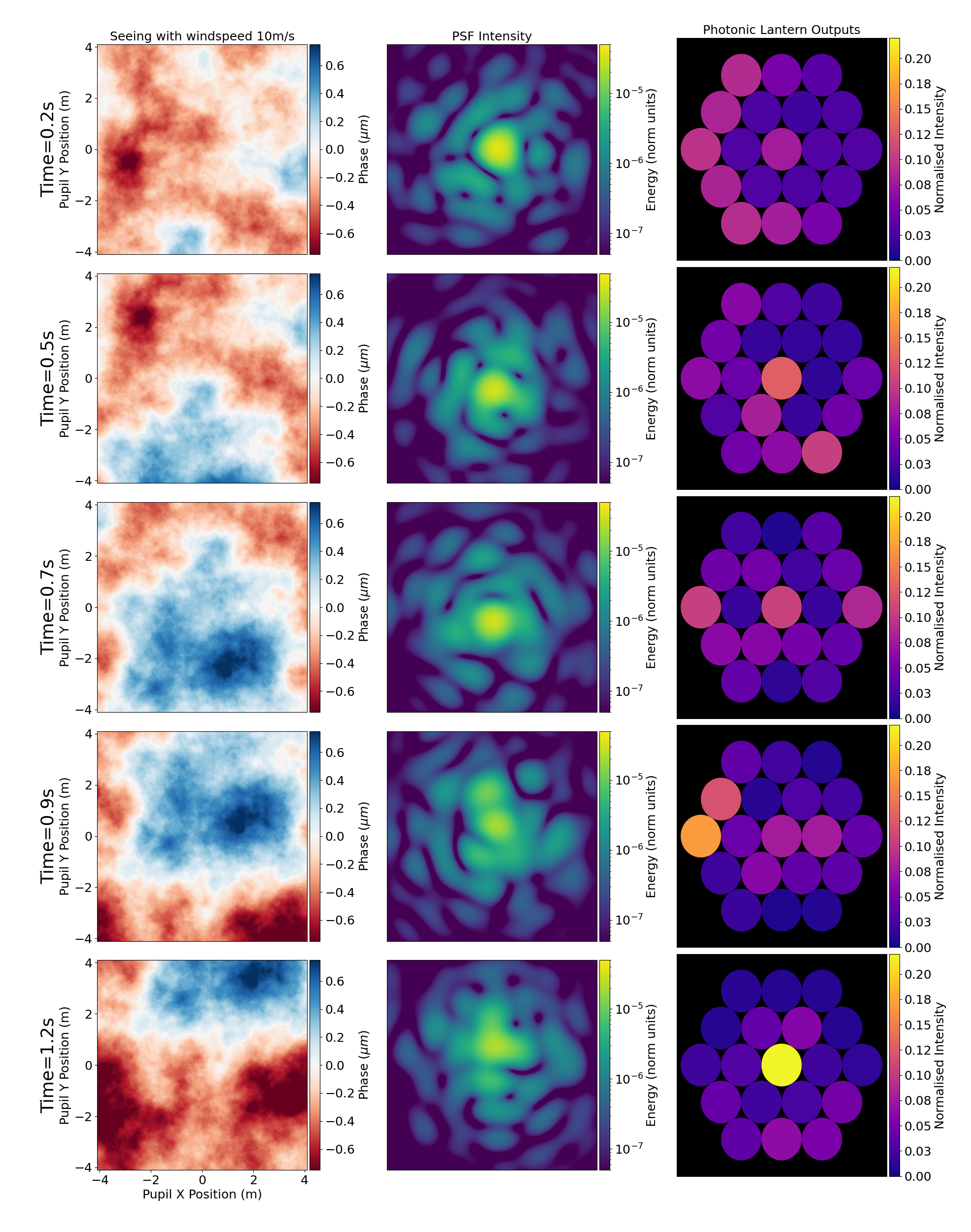}
    \caption[`Film-strip' style depiction of video showing neural-network simulation results wherein seeing is passed through a photonic lantern]{A `film-strip' style representation depicting a segment of video showing neural-network simulation results, wherein 30 seconds of seeing is propagating through a PL. The left-hand panel shows the turbulent phase screen which was generated for a windspeed of $10$~$m/s$, the middle panel shows the PSF of the system and the right-hand panel shows a depiction of the output illumination of the PL. The seeing was generated with a Fried parameter of $0.7$~$m$ for a telescope with a diameter of $8.2$~$m$. The intensity was normalised such that the intensity across all waveguides for a flat wavefront summed to 1. (Video 1, MP4, 11.9 MB)}
    \label{fig:seeing_video}
\end{figure}

\begin{figure}[htb!]
    \centering
    \includegraphics[width=0.4\textwidth]{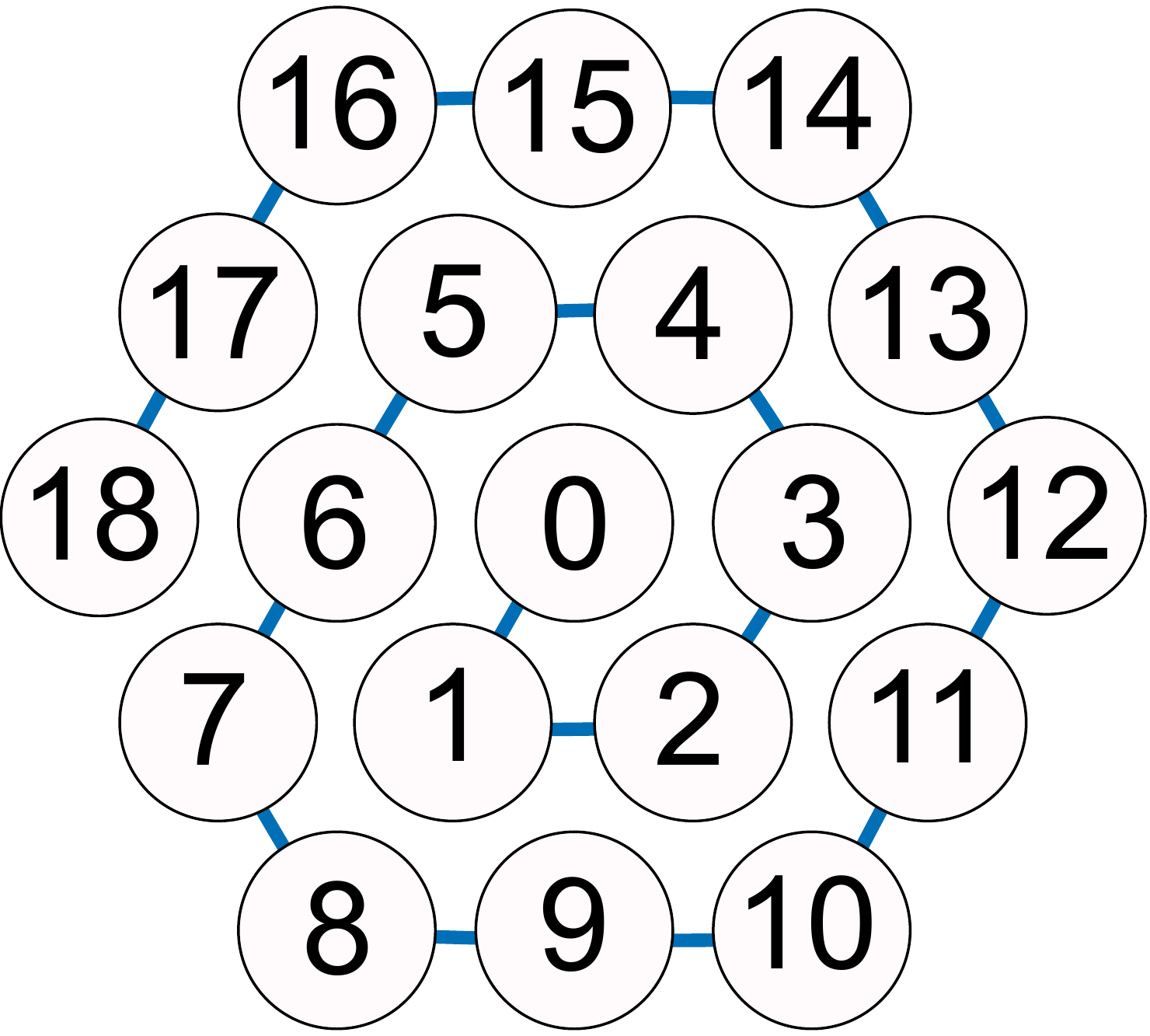}
    \caption[Numbering of waveguides]{Diagram showing the numbering system adopted of the single-mode waveguides at output.}
    \label{fig:waveguide_pattern}
\end{figure}

\begin{figure}[htb!]
    \centering
    \includegraphics[width=0.6\linewidth]{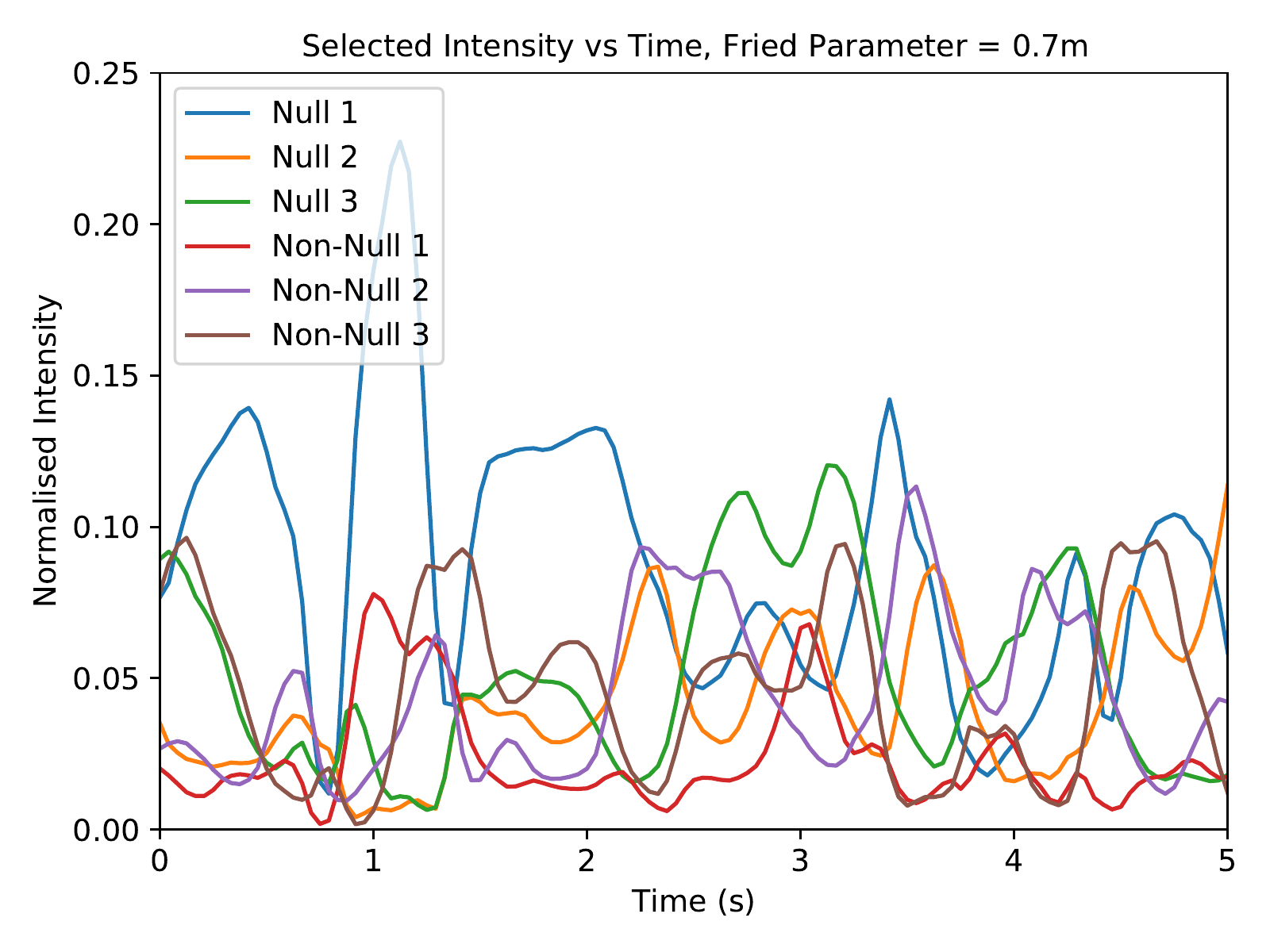}
    \caption[Intensity versus time through selected waveguides as seeing is passed through a photonic lantern]{A plot of neural-network simulation results showing the intensity of selected waveguides versus time, as seeing is passed into the PL. Waveguide 0 is the central waveguide, waveguide 1 is a waveguide in the inner-ring, waveguide 8 is a {corner}-waveguide in the outer ring and waveguide 9 is an {edge}-waveguide in the outer ring. This seeing was generated with a a windspeed of $10$~$m/s$ and a Fried parameter of $0.7$~$m$ for a telescope with diameter of $8.2$~$m$. The intensity was normalised such that the intensity across all waveguides for a flat wavefront summed to 1. (Video 1, MP4, 11.9 MB)}
    \label{fig:PL_line_plot}
\end{figure}

\begin{figure}[htb!]
    \centering
    \includegraphics[width=0.6\linewidth]{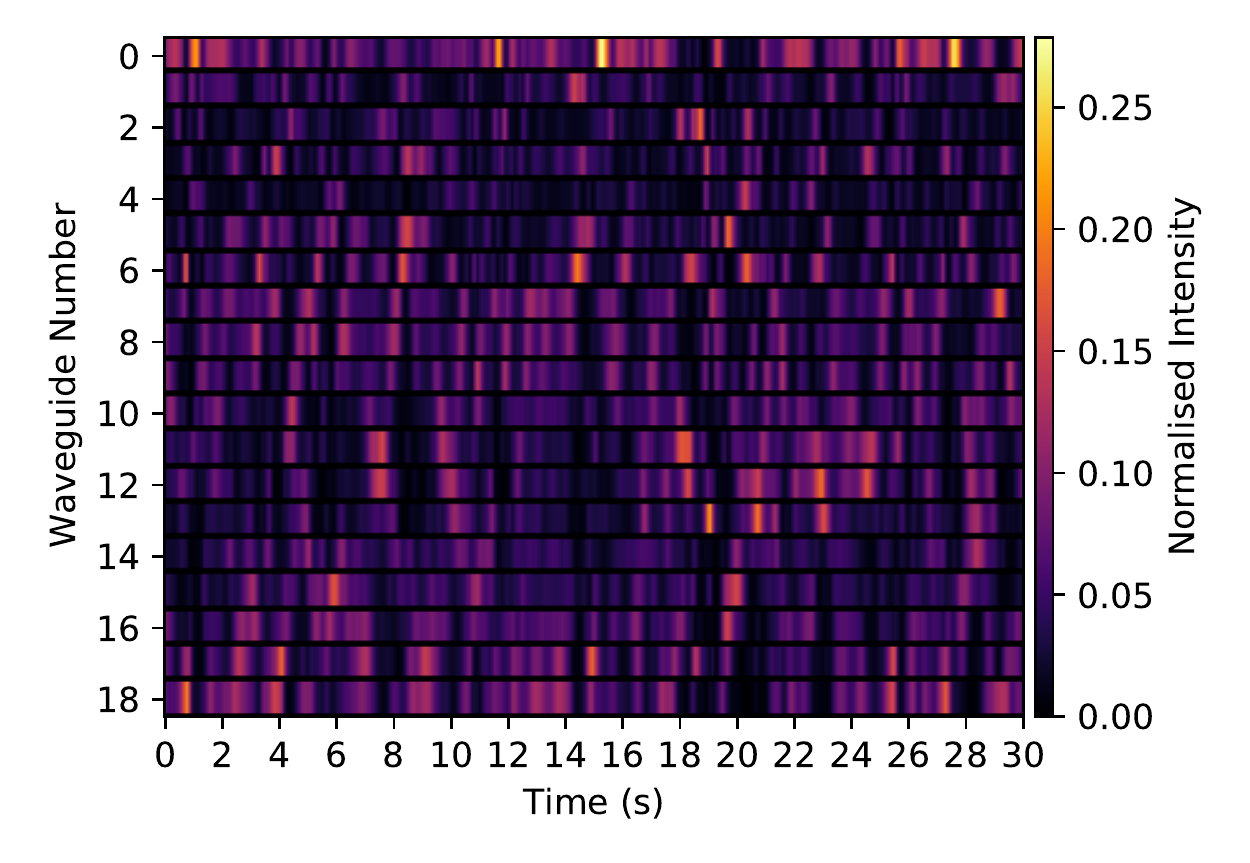}
    \caption[Intensity versus time through selected waveguides as seeing is passed through a photonic lantern]{A waterfall-style plot of neural-network simulation results, showing the shifting waveguide intensities in response to seeing. Seeing was generated assuming a windspeed of $10$~$m/s$ and a Fried parameter of $0.7$~$m$ for a telescope with diameter of $8.2$~$m$. The intensity was normalised such that the intensity across all waveguides for a flat wavefront summed to 1. (Video 1, MP4, 11.9 MB)}
    \label{fig:waveguide_waterfall}
\end{figure}

\begin{figure}[htb!]
    \centering
    \includegraphics[width=0.8\linewidth]{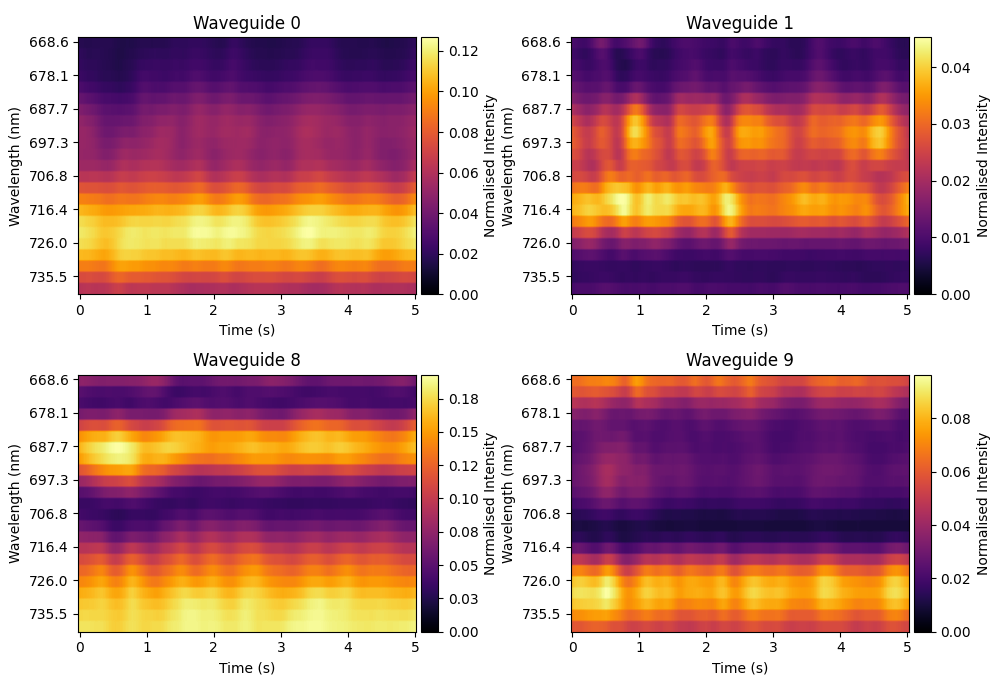}
    \caption[Polychromatic intensity versus wavelength through time as seeing is passed through a photonic lantern]{A waterfall-style plot of polychromatic neural-network simulation results, showing the intensity as a function of wavelength as seeing is passed through the NN model, for selected waveguides. Waveguide 0 is the central waveguide, waveguide 1 is a waveguide in the inner-ring, waveguide 8 is an edge-waveguide in the outer ring and waveguide 9 is a corner-waveguide in the outer ring. In sections where the seeing is outside the range of the model (i.e. one of the Zernike coefficients is greater than $\sim 0.35$ radians) predictions may be less reliable. The intensity was normalised such that the intensity across all waveguides for a flat wavefront summed to 1.}
    \label{fig:broadband_seeing}
\end{figure}

In order to simulate aberrated wavefronts distorted by seeing which are then passed into a PL, we used HCIPy \cite{Por2018HighSimulator} to generate realistic seeing phase screens, and then OpticsPy \cite{Fan2019OpticsPy} to decompose the wavefronts over the Zernike polynomials (as the NN models were trained on this basis). Thirty seconds of numerically-generated seeing data, sampled at $40$~ms time-steps, was fed into the best-performing NN model on the 8 Zernike data set. The results, presented as a multi-frame ``video'', can be seen in Fig.~\ref{fig:seeing_video}{. }A plot illustrating the intensities of selected waveguides---numbered in Fig.~\ref{fig:waveguide_pattern}---as a function of time can be seen in Fig.~\ref{fig:PL_line_plot} or as a waterfall plot in Fig.~\ref{fig:waveguide_waterfall}. 
With the polychromatic model, trained on a physical PL, we are also able to see the relationship between wavelength and mode excitement as seeing is passed into the lantern, as shown in Fig.~\ref{fig:broadband_seeing}.

{}

{}

{}

{}

{}

Passing each frame of the video to a traditional beam propagation algorithm would have taken over a month; by using the NN model it took around 30 seconds. The models are by no means limited to 30 seconds of input; they can compute the PL outputs for as many wavefronts as are provided for $\sim0.03$~$s$ per wavefront. 


\subsubsection{Funnels and Nullers}

When attempting to inject light from an AO-corrected telescope into a fiber or PL, the usual approach is to maximise the Strehl ratio of the beam (as is usual in AO operations) in order to maximise coupling \cite{NemSMF, NemStrehl}. Sometimes apodiziation, such as phase-induced amplitude apodization\cite{PIAA, NemStrehl, Calvin:21}, is used to further improve coupling. However, other potential applications exist wherein a specific distribution of intensities in the output waveguides from a PL is desired. To achieve this (especially in the case of a real, manufactured PL), an incident wavefront must be identified which optimally produces the desired output intensities. This wavefront becomes the new zero-point of the AO system (rather than aiming for the flattest possible wavefront), and so is restricted to the space of wavefronts that can be produced by the DM. This is defined by the actuator number, spacing and maximum stroke {of} the DM, and the given pupil amplitude distribution.
Determining this wavefront thus requires a global optimisation over this space, with some potentially complicated or degenerate merit function.


One such application is a so-called ``PL funnel''. 
This is a {system} that translates some low Strehl incident optical state and injects it, with high efficiency, into just one (or a small number of) single-mode waveguides emerging from a PL---rather than distributing it between all outputs. 
{One could alternatively aim to just directly inject the light into one single-mode fibre, but this has proved extremely challenging due to the high-degree of wavefront correction required (hampered by non-common path errors between fibre and wavefront sensor) and mismatch of telescope pupil amplitude (and resulting PSF) with the fibre's mode field. By using a PL the target wavefront can be optimised to maximise injection, and light leaking into the `dark' outputs could be used as a wavefront sensor to maintain this target wavefront without non-common path errors} \cite{Norris2020AnSensor}{. It also allows light to be injected into a small number of waveguides, capturing light which would be lost (due to poor AO correction) if a single single-mode fibre was used.
Once light is in a single-mode fibre}, the size of a spectrograph no longer scales with the number spatial modes in a telescope, allowing small, compact spectrographs\cite{Bland-Hawthorn2010PIMMS:Microspectrograph, Betters:13, Schwab2012SingleSpectrographs, 8231484, Jovanovic_2016, iLocater}. These small spectrographs are more easily stabilised and cheaper to fabricate, enabling a greater number of objects to be observed concurrently if several are placed in a telescope. 

A PL nuller is the converse of an PL funnel. 
Instead of attempting to identify a wavefront which {\em maximises} the coupling into a single waveguide of a PL we seek to {\em minimise} light into that waveguide\cite{Norris2019FirstInstrument}. 
A nulling interferometer (hereafter nuller) is particularly useful when attempting to observe faint objects (such as an exoplanet) which are in close proximity to a bright source (typically the host star) \cite{Bracewell1978DetectingInterferometer}. 
Interference is created between two telescopes (or regions of a telescope pupil) to cancel out the unwanted bright source, such as a central star,  preserving the much fainter signal from nearby objects (e.g. circumstellar disk or planet). This has previously been done using traditional bulk optics\cite{Colavita:09, Defrere:16} and also by photonic techniques in testbed instruments such as GLINT\cite{Norris2019FirstInstrument} and PFN\cite{Mennesson:11, Kuhn:15}. 
As explored here, by carefully choosing an incident wavefront a PL could potentially serve a similar function.

Identifying the precise wavefronts that result in funnels or nullers can be accomplished with a global optimsation algorithm. For this project, SciPy's \cite{Virtanen2020SciPyPython} \texttt{basinhopping} optimisation function was employed which implements the basin-hopping algorithm described by Wales \& Doye\cite{Wales1997GlobalAtoms}. These optimisations explored the space of Zernike polynomial coefficients which, when passed through the NN model, achieved waveguide intensities that minimised our objective function (see Equation~\ref{eq:null_equation} below).

Over the course of the optimisation process described here, around 65\,000 calls to the model were made. 
This would have taken of order 5 and a half years if the wavefronts had been passed through a traditional beam propagation algorithm.
Using the models constructed for this work (which take around 0.03 seconds to map an incident wavefront into a pattern of single-mode output intensities) the entire global optimisation described can run to completion in under an hour.

In this work, the most severe of the constraints are (1) the restriction on the amplitude (constant across the pupil, apart from zeros at regions shadowed by telescope structures), and (2) the limited set of spatial frequencies that can be produced by the DM.
Here, the set of functions chosen to represent the phase is the Zernike basis.
The goal is to maximise (or minimise) the light through a number of waveguides while remaining agnostic to which specific waveguides contribute.

\begin{figure}[htb!]
    \centering
    \includegraphics[width=1\textwidth]{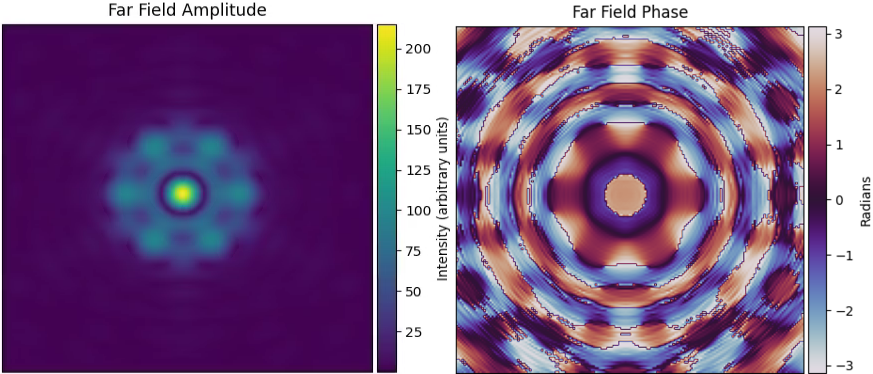}
    \caption[Wavefront required for perfect optical funnelling through central waveguide]{Depictions of the phase and amplitude required to create a perfect PL funnel into the central waveguide, i.e. a wavefront which results in 100\% of light passing through the central waveguide. Note the non-uniform amplitude in the far field {(equivalent to the telescope pupil)} and very high spatial frequencies---which prevents the formulation of this wavefront by a DM.}
    \label{fig:reverse_propagation}
\end{figure}

If no such constraints existed, and there was no degeneracy in the desired output intensity pattern (e.g. all flux in one specific waveguide), then it is trivial to obtain an ideal solution for a theoretical PL.
Employing traditional beam propagation software such as RSoft, propagating light from the single desired waveguide in the reverse fashion through the lantern yields the electric field at the multimode region before the adiabatic taper. 
Conjugating this electric field results in a wavefront which, if passed through the PL, would maximally excite that single waveguide. 
This method yielded the solution shown in Fig.~\ref{fig:reverse_propagation}, which, it was confirmed, resulted in the expected perfect funnel operation under forward propagation back through the lantern.
Unlike situations where two or more waveguides are illuminated, in the case of a single illuminated waveguide the solution is uniquely defined.
Unfortunately, as the wavefront is described by strong, high-frequency variations in both amplitude and phase, it violates the terms of our constraints listed above.
Therefore the task becomes an optimisation problem. 

{}

One example is to optimise for the largest possible fraction of the light being contained in any one single waveguide, with the remaining light shared in any fashion among the others. This allows the optimiser to choose not only the optimum achievable wavefront, but also the target output waveguide which allows the best results. 
A second possibility is to identify an input which maximises the total light channeled into some small number of waveguides; for example, attempting to find a pattern yielding two or three waveguides containing almost all the flux. 

Unlike a perfect PL funnel, the solution for a PL nuller is not unique. 
Therefore solutions which completely extinguish the output of a single waveguide are relatively plentiful. 
Additionally, we may choose to add further (scientifically valuable) constraints; specifically, nulling multiple waveguides simultaneously. Indeed, the optimisation can find such solutions allowing us to further explore the trade-off between the number of near-null outputs and the degree of extinction (quantified by, for example, the cumulative light into the nulled waveguides).

\subsubsection{Identified funnel and nuller solutions}
\label{sec:funnels_and_nullers}

\begin{figure}[htb!]
    \centering
    \includegraphics[width=0.8\textwidth]{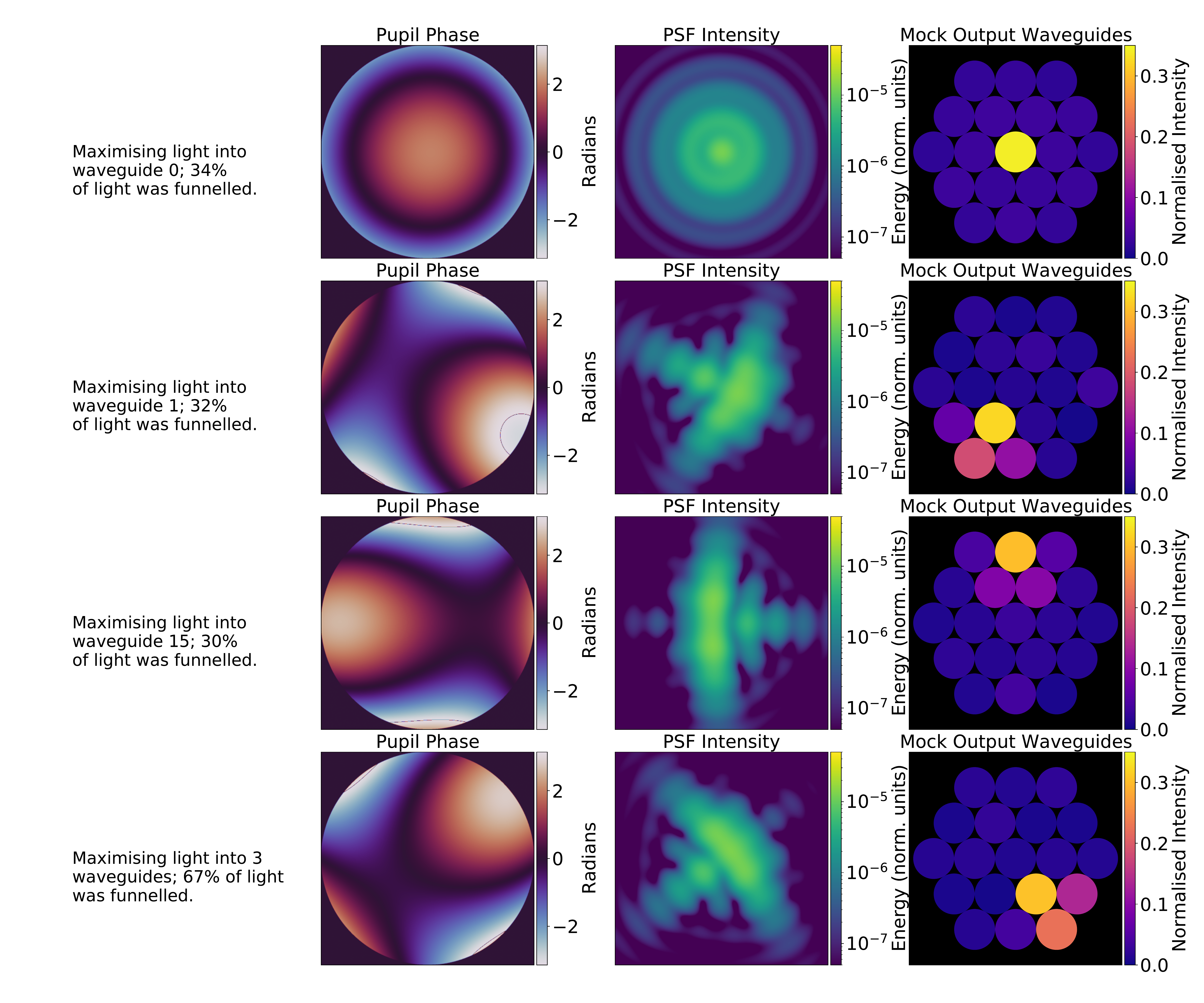}
    \caption[A selection of photonic lantern funnel results]{Depiction of the results of the basin-hopping algorithm utilising the 8-Zernike $3.15$~radian RMS WFE model, tasked with seeking to maximise the light intensity through each individual waveguide. {Note that since the model only supports 8 Zernike polynomials the maximum spatial frequency is limited by this constraint rather than the spatial-light modulator; using a model with a greater number of supported Zernike polynomials would expand this limit.} The left image in each row is a representation of the phase of the wavefront at the pupil, i.e. as it arrives at the optical system, the central image is the PSF which is injected into the PL, and the final image is an end-on depiction of the PL showing the intensity of each waveguide. The intensity was normalised such that the intensity across all waveguides for a flat wavefront summed to 1.}
    \label{fig:single_funnels}
\end{figure}

As our lantern contains 19 waveguides at output and we are restricted to modulating pupil-plane phase, there are 19 solutions which maximise light into a single waveguide.
These were found using the basin-hopping algorithm, with the objective function defined to minimise the flux in a given waveguide. 
Figure~\ref{fig:single_funnels} shows some of the outcomes produced by these optimisations using the 8-Zernike $3.15$~radian RMS WFE model.

These results show that the best funnel for a single waveguide attempts to maximise light in the central waveguide, achieving 34\% of total intensity. Solutions for the best funnel for the 1st, 2nd, 3rd, 4th, 5th, and 6th waveguides---numbered in Fig.~\ref{fig:waveguide_pattern}---are all hexagonal rotations of the same underlying pattern, as expected. Similarly the same degeneracy exists in solutions among the 7th, 9th, 11th, 13th, 15th and 17th and the 8th, 10th, 12th, 14th, 16th and 18th waveguides due to the six-fold rotational symmetry in the system as was discussed in the data augmentation portion of Sec.~\ref{sec:simulated_data}.

{}

\begin{figure}[htb!]
    \centering
    \includegraphics[width=0.75\textwidth]{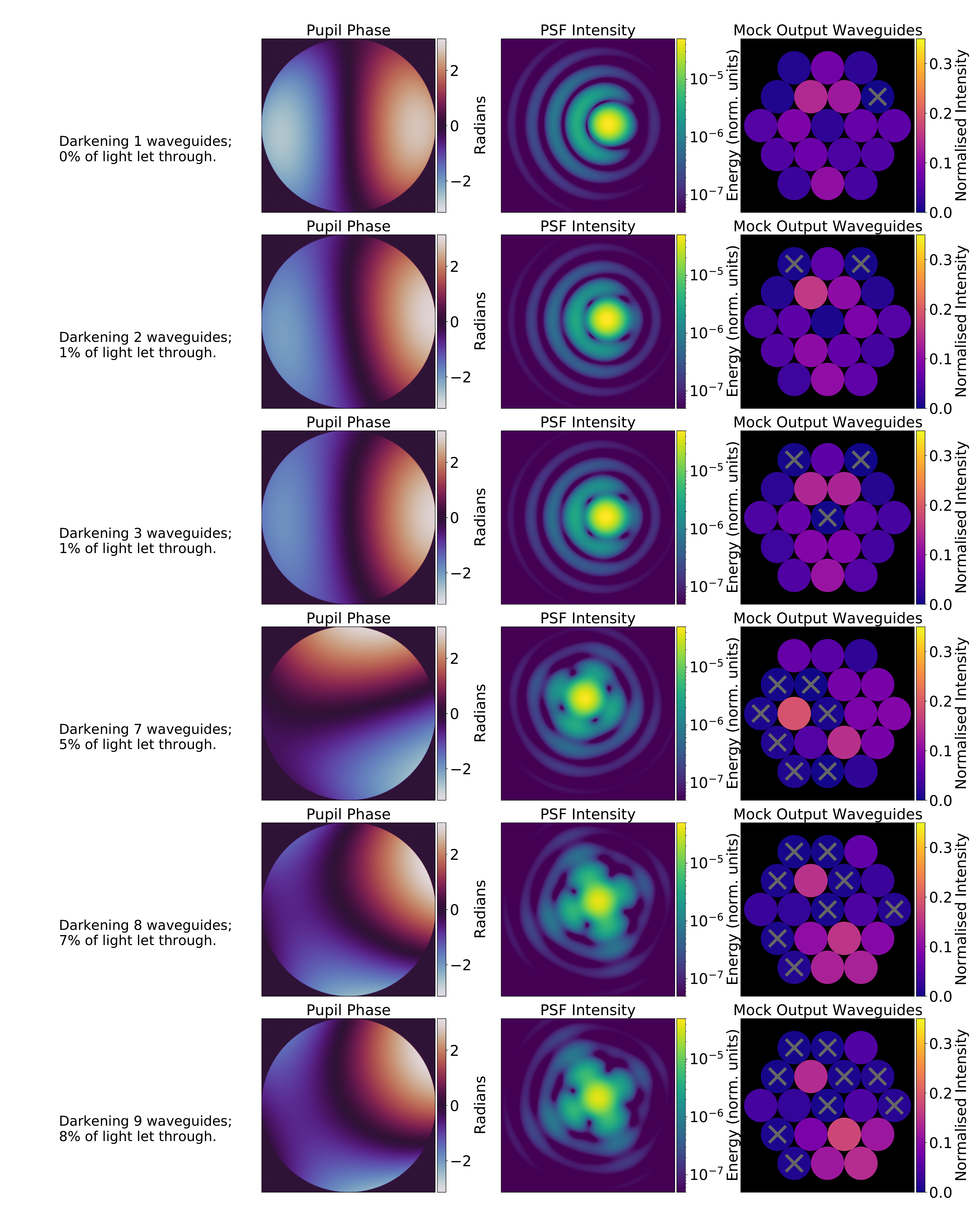}
    \caption[A selection of optical nulling results]{Depiction of the results {(accurate to the nearest percent)} of the basin-hopping algorithm utilising the best performing 8-Zernike $3.15$~radian RMS WFE model, tasked to minimise the light through some number of waveguides. {The crosses indicate the waveguides which were chosen by the algorithm to darken. The robustness of these nulls to an incorrect DM position is explored in Fig.~}\ref{fig:seeing_with_nuller}. The left image in each row is a representation of the phase of the wavefront at the pupil, i.e. as it arrives at the optical system, the central image is the PSF which is injected into the PL in each case and the final image is an end-on depiction of the PL showing the intensity of each waveguide after the lantern has been injected with this wavefront. The intensity was normalised such that the intensity across all waveguides for a flat wavefront summed to 1.}
    \label{fig:optical_nullers}
\end{figure}

A similar technique can be used to null a set of outputs. Figure~\ref{fig:optical_nullers} shows the solutions found by the basin-hopping algorithm set to extinguish the exiting light through the specified number of waveguides---note that we don't specify which waveguides---while also maximising the total amount of light which propagates through the lantern.
This was done by combining these two goals into a single objective function as is shown in Eq.~(\ref{eq:null_equation}) below, which was then set to be minimised by the gradient descent algorithm:
\begin{align}
    f(x, n) &= 1 - \sum_{i=0} ^{18} x_i + \sum_{i=0} ^n x_i \label{eq:null_equation}
\end{align}
Where $x$ is a list of waveguide intensities sorted in ascending order, $x_i$ is the $i$th lowest intensity of the waveguides and $n$ is the number of waveguides to be nulled. Note that there are 19 waveguides in the PL used in this work and so the sum from $x_0$ to $x_{18}$ is a sum over every waveguide.

{}

These results show that a wavefront solution for a PL nuller capable of effectively extinguishing a single waveguide, or even a handful of waveguides, can be recovered. 
If we wish to extinguish larger numbers of waveguides, then we must accept solutions which let an increasing fraction of light through the nulled waveguides.
As shown in Fig.~\ref{fig:optical_nullers}, 1--3 nulled waveguides let 0--1\% {(to 1 significant figure)} of light through, attempting nulling on 7 waveguides results in $\sim 5\%$ of light leakage into those guides, while a 9-waveguide design results in $\sim 8\%$ leakage into the nulled channels. {Numbers stated here are accurate to the nearest percent, since the MSE of this model is $1.22\times 10^{-4}$~radians the RMS error is of order $10^{-2}$.}

\begin{figure}[htb!]
    \centering
    \includegraphics[width=1\textwidth]{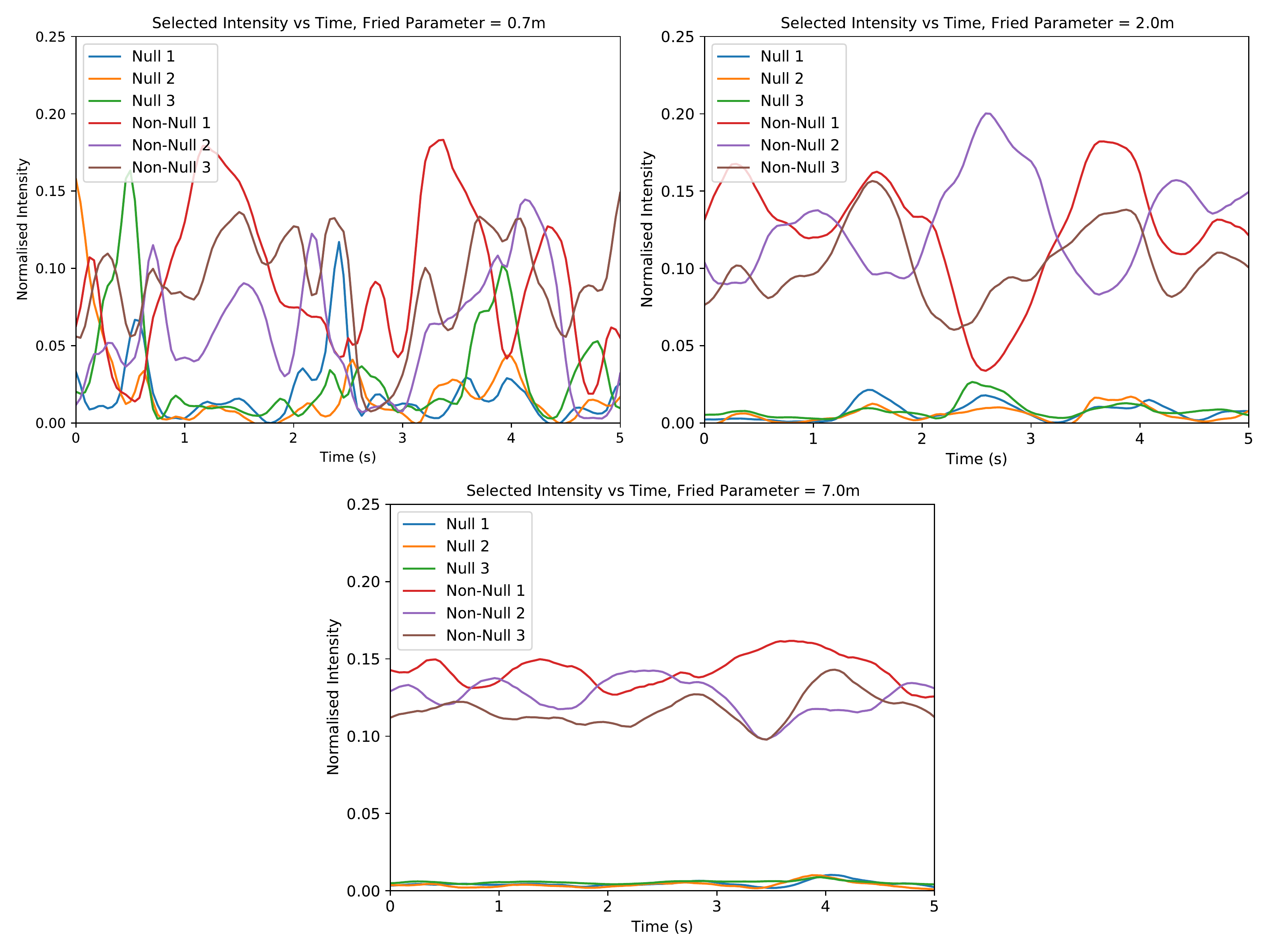}
    \caption[Nuller applied to seeing]{Diagram showing the effect of applying our 3-nuller solution to seeing. The seeing in the top-left plot was created with a Fried parameter of $0.7$~$m$. The seeing in the top-right plot was created with a Fried parameter of $2$~$m$. The seeing in the bottom plot was created with a Fried parameter of $7$~$m$. In all cases the data was created for telescope with an $8.2$~$m$ diameter experiencing a wind speed of $10$~$m/s$. The intensity was normalised such that the intensity across all waveguides for a flat wavefront summed to 1.}
    \label{fig:seeing_with_nuller}
\end{figure}

{We can then examine the effect of these solutions when encountering atmospheric seeing (or residual WFE after AO correction). The seeing was generated as described in Sec.}~\ref{sec:seeing_response}. {The found solution wavefront was effectively set as the new zero-point of the AO system by adding it to the seeing-induced wavefront, and so the nulled waveguides, assuming good seeing correction, should be extinguished. A line plot showing the results of this nulling can be seen in Fig.}~\ref{fig:seeing_with_nuller}. It shows that as the seeing is better corrected, the PL nuller is better able to null out the specified waveguides.

{}

\section{Future Work}
Training a NN model capable of making predictions based on wavefronts described by more Zernike polynomials would simply require more data. While the techniques outlined in Sec.~\ref{sec:simulated_data} mitigate the need for data (and make it faster to generate), it is still ultimately a stumbling block for creating these NN models from beam-propagation simulations. We found that as the size of the Zernike space increased so too did the number of hidden layers which resulted in the best performance. Thus, increasing the number of Zernike polynomials or the RMS WFE range will likely involve training larger NNs. As the number of hidden layers increases it is possible that a ResNet-style structure\cite{He2016DeepRecognition} will prove optimal. We found that larger models were best regularised with Dropout.

We chose to use the Zernike polynomials to describe our input wavefront however this is not the only choice; future NN models could use a {actuator} basis of the {DM}. While this greatly increases the dimensionality of the input space it does remove some of the restrictions imposed on the space by the use of a finite number of Zernike polynomials. {There is promising research suggesting that a} CNN\cite{NIPS1989_53c3bce6} {would be best suited to this task due to its innate spatial reasoning}.

While we demonstrated NN model performance on a single physical PL (see Table~\ref{table:models_practical}) there is no reason these models shouldn't work on any PLs if training data is provided. 

With enough training data it should be possible to characterise propagation through a PL with a physical structure unseen in the training set. Similar to the polychromatic network structure described in Sec.~\ref{sec:lab_data}, a network could be trained which took as inputs various parameters which describe the PL's design. Once trained, such a model could then predict the outputs of PLs with arbitrary properties. This would allow PLs which vary in waveguide length, waveguide width, waveguide refractive index or cladding refractive index to be characterised. The volume of data required for this characterisation is estimated to be 1--2 orders of magnitude greater than that for a single PL.

\section{Conclusion}


Photonic lanterns are an exciting technology with a wide variety of applications in astronomy, optics and telecommunications. Successful, widespread deployment for single-mode spectroscopy, interferometry and other astrophotonic applications will require fast and accurate simulation tools for design and testing. However, traditional beam propagation algorithms applied to these three dimensional photonic devices are {slow enough} ($\sim 1$ hour per propagation on a single {quad-core} CPU) to make them impractical for some uses. We developed neural network models capable of accurately predicting a photonic lantern's output in $0.03$~$s$, a speed-up of over 5 orders of magnitude. This technique was used to model existing photonic lanterns---including manufacturing defects---and was also shown to generalise to polychromatic inputs. Using these models, the response of each of the photonic lantern's outputs can be predicted as a function of an arbitrary seeing time-series, over a range of wavelengths. We also demonstrated two possible applications: modelling a photonic lantern's response to seeing and using a global optimiser to identify wavefront shapes which act as photonic lantern funnels and nullers.

\subsection* {Acknowledgments}

We would like to thank the team at Olivier Guyon, Nemanja Jovanovic and Julien Lozi for their stimulating discussions.

\subsection* {Code, Data, and Materials Availability} 

We have made all code freely available at: \\ \url{https://github.com/David-Sweeney/Learning-the-Lantern}

Simulated data is freely available in the supplementary material. Data from the optical testbed may be made available upon personal request.

\bibliography{report}   
\bibliographystyle{spiejour}   


\vspace{2ex}\noindent\textbf{David Sweeney} is a PhD student at the University of Sydney. He completed a Bachelor of Science (Advanced) (Honours) in 2020 with First Class Honours in Physics and a major in Computer Science. His research interests centre on the application of machine learning techniques to problems in other fields.

\vspace{2ex}\noindent\textbf{Richard Scalzo} is a Senior Research Fellow at the ARC Industrial Transformation Training Centre in Data Analytics for Resources and the Environment (DARE).  His research focuses on the fusion of probabilistic and deterministic models in applications across the physical sciences, including astrophysics, geophysics, geology, and hydrology.

\vspace{1ex}
\noindent Biographies and photographs of other authors are not available.


\end{spacing}
\end{document}